\documentclass[reqno]{amsart}

\usepackage[margin=3cm]{geometry}

\usepackage{arxiv_style}
\usepackage{fontawesome}

\makeatletter
\def\blfootnote{\gdef\@thefnmark{}\@footnotetext}
\makeatother

\title[TENET: Telegram Mini App (in)security]{TENET: Telegram Mini App (in)security}

\author[A.\ Ciccotelli, F.\ Zappone, and R.\ Di Pietro]{%
    Andrea Ciccotelli\textsuperscript{1}\,\orcidlink{0009-0004-8286-5135},
    Federico Zappone\textsuperscript{2}\,\orcidlink{0000-0001-6455-6575}, and
    Roberto Di Pietro\textsuperscript{1}\,\orcidlink{0000-0003-1909-0336}%
}

\dedicatory{%
    \normalfont
    \begin{minipage}[t]{0.47\linewidth}
        \centering
        \textsuperscript{1}\textbf{King Abdullah University of Science and Technology (KAUST)}\\
        Computer, Electrical and Mathematical Sciences and Engineering (CEMSE) Division\\
        Thuwal 23955, Saudi Arabia\\
        \href{mailto:andrea.ciccotelli@kaust.edu.sa}{andrea.ciccotelli@kaust.edu.sa}\\
        \href{mailto:roberto.dipietro@kaust.edu.sa}{roberto.dipietro@kaust.edu.sa}
    \end{minipage}\hfill
    \begin{minipage}[t]{0.47\linewidth}
        \centering
        \textsuperscript{2}\textbf{Artificialy}\\
        Lugano, Switzerland\\
        \href{mailto:federico.zappone@artificialy.com}{federico.zappone@artificialy.com}
    \end{minipage}\\[1em]%
}

\begin{document}

\begin{abstract}
Telegram, with over 450 million daily active users, has introduced Mini Apps---web-based applications running directly within its client. However, this integration introduces notable security risks. As we demonstrate, many Mini Apps store authentication materials---such as session tokens and wallet mnemonic phrases---in plaintext on client devices, exposing users to unauthorized access, impersonation, and financial exploitation. While insecure client-side storage is a known risk in web applications, the Telegram Mini App ecosystem presents a uniquely dangerous combination of factors absent from prior work: no platform-level security review, no storage access restrictions, a financially motivated user base handling live cryptocurrency assets, and a WebView environment that offers weaker protections than standalone browsers. To investigate this threat, we present TENET, a purpose-built auditing tool whose design decisions---pattern selection, entropy thresholds, and charset validation---are grounded in the structural properties of the secrets targeted and empirically validated against a ground-truth dataset. We screened 61 Mini Apps using a stratified, popularity-weighted sampling strategy. Of the 37 applications that met our processing criteria and were analyzed, 30 exhibited security flaws, which we classify into three severity tiers: plaintext storage, recoverable encryption, and replayable tokens. Notably, even Telegram's official Wallet exhibits a severe vulnerability that may lead to full account compromise. Following our responsible disclosure, Telegram implemented two new secure-storage APIs, and our post-remediation verification confirmed that its official Wallet no longer exposes the recovery mnemonic in plaintext. Finally, we propose mitigation measures and best practices for both Telegram platform developers and third-party Mini App creators.
\end{abstract}

\maketitle
\thispagestyle{empty}

\begin{multicols}{2}
    % \small
    % \vspace{-1em}
    \tableofcontents
    % \vspace{-3.5em}
\end{multicols}

% \newpage
\section{Introduction}
\label{sect:intro}

Since its release in 2013, the cloud-based messaging and communication platform Telegram~\cite{telegram} has evolved far beyond a simple messaging service to become a comprehensive ecosystem that hosts a diverse range of integrated applications, known as \emph{Mini Apps}~\cite{telegram}. These applications are built using standard web technologies (HTML, JavaScript, and CSS) and are embedded directly within the Telegram client. They leverage a controlled set of APIs to offer functionalities that range from basic information retrieval to complex financial transactions while maintaining a seamless user experience within a single unified interface.

Integrating Mini Apps into Telegram has enabled external developers to innovate rapidly. However, this architectural convergence of native and web-based components gives rise to a complex security landscape.
\HIGHLIGHT{Telegram protects ordinary Cloud Chats through client--server encryption, while end-to-end encryption is available for Secret Chats and calls; the local storage mechanisms used by Mini Apps nevertheless remain dependent on third-party developers' security practices.} In many cases, sensitive information---including personal identifiers and financial credentials like wallet mnemonic phrases---is stored in plaintext or using weak encryption schemes. The decentralized nature of Mini App development, where best practices~\cite{kandula2024webview} are not consistently followed, makes the ecosystem a lucrative target for attackers. Similar phenomena have been observed in other modular ecosystems~\cite{10.1007/978-3-032-07894-0_20,mallojula2024companion}. Given that the TON blockchain ecosystem alone processes billions of dollars in transactions~\cite{ton_transactions} and Telegram's native Wallet has surpassed 100 million activations~\cite{TgramWall}, the financial exposure is substantial.

In this paper, we conduct the first systematic measurement study of client-side storage security in the Telegram Mini App security ecosystem, with a focus on applications that transact on crypto tokens---the primary target of financially motivated adversaries.
We analyze the methodologies used for local data storage, identify critical vulnerabilities that allow sensitive data to be easily extracted, and demonstrate practical exploitation scenarios.
Our analysis proposes a data extraction and analysis technique that operates on the victim's local device. We emphasize that the local-access assumption is realistic in the context of infostealers and commodity malware (e.g., RedLine, Raccoon, Vidar), which routinely harvest LevelDB and SQLite stores from user space without root privileges. \HIGHLIGHT{We additionally demonstrate two non-physical-access scenarios: an XSS flaw in a vulnerable Mini App can exfiltrate secrets stored under that Mini App's origin, while a malicious document executed with user-level privileges can copy Telegram's storage directory without requiring physical access.}
Furthermore, we propose viable countermeasures that can be implemented at the Telegram platform level or by individual Mini App developers to mitigate these risks significantly.

\textbf{Contributions:}
Our analysis of Telegram Mini Apps highlights the urgent need for rethinking the security of such applications.
Indeed, our findings show that Personally Identifiable Information (PII) and financial information can be easily extracted from a victim's device once user-space access is obtained.
Our contribution can be summarized as follows:
\begin{itemize}
    \item \HIGHLIGHT{We present the first systematic, large-scale measurement study of client-side storage security in the Telegram Mini App ecosystem, screening 61 apps selected via a stratified popularity-weighted sampling strategy and successfully analyzing 37 of them;}
    \item We introduce a three-tier vulnerability taxonomy (plaintext storage, recoverable encryption, replayable tokens) that enables severity-aware assessment of insecure Mini App storage;
    \item We propose techniques for extracting sensitive information from Telegram Mini App storage in user space, without requiring root privileges;
    \item We develop TENET---a purpose-built auditing tool whose 27 detection patterns, entropy thresholds, and charset validators are each justified by the structural properties of the targeted secret class, and whose performance is empirically evaluated \HIGHLIGHT{in terms of precision, recall, F1-score, and runtime}. TENET will be released under an open-source license;
    \item We provide a list of vulnerable cryptocurrency-based Mini Apps, and we highlight what sensitive information we found;
    \item We propose a solution to this vulnerability implementable at the Telegram or Telegram Mini App levels; and,
    \item \HIGHLIGHT{Following our responsible disclosure, Telegram introduced two new secure-storage APIs and remediated the official Wallet vulnerability.}
\end{itemize}

%%---------------------------------------------------------------------
\section{Related Work}
\label{sect:related_work}

Early research on Telegram's security focused on its proprietary MTProto protocol. Sušánka and Kokeš~\cite{10.1145/3150376.3150382} identified fundamental issues such as undocumented obfuscation techniques and replay attack vulnerabilities. Abu-Salma et al.~\cite{abu_salma} found that confusing interface designs led users to misconfigure security settings. Albrecht et al.~\cite{9833666} uncovered timing side-channel and replay attacks in the official client, while Von Arx and Paterson~\cite{10.1145/3579856.3582811} extended these findings to third-party implementations. Maréchal~\cite{220215} examined how Telegram's security model interacts with censorship and surveillance, and La Morgia et al.~\cite{10248275} explored the misuse of Telegram channels for misinformation and fraud.

While these studies advanced the understanding of Telegram's security, they overlooked client-side data management within Mini Apps. These applications integrate web-based functionalities directly into the native client and rely on standard web storage mechanisms---such as \texttt{localStorage}, \texttt{sessionStorage}, cookies, and IndexedDB---to cache data, introducing a new threat surface that traditional analyses of Telegram's messaging protocol do not cover. In the broader context of web applications, Hanna et al.~\cite{hanna2010emperor} examined persistent client-side XSS vulnerabilities exploiting \texttt{localStorage}. Zheng et al.~\cite{zheng2015cookies} investigated risks in cookie storage, revealing that improperly secured cookies can be exploited to hijack user sessions. Steffens et al.~\cite{steffens2019dont} showed that many SPAs fail to adequately sanitize data stored in \texttt{localStorage} and \texttt{sessionStorage}, underscoring the need for robust client-side storage encryption.

Recent studies on mini-application ecosystems confirm that inadequate data-handling is not unique to Telegram. MiniTracker~\cite{wei2024minitracker} detects large-scale leaks in WeChat mini-apps. Jun Li et al.~\cite{10945038} analyse permission-control flaws, Shuai Li et al.~\cite{10409279} uncover cross-user privacy leakages, and Yue Zhang et al.~\cite{10.1145/3576915.3616591} demonstrate the impact of leaked \texttt{AppSecret} keys in WeChat mini-programs. However, these works target ecosystems (primarily WeChat) that enforce platform-level sandboxing, mandatory API vetting, and developer certification---mechanisms that are absent in Telegram. Unlike WeChat, Telegram imposes no security review on Mini App submissions, does not restrict local storage access, and delegates all data-protection responsibilities to third-party developers---making a dedicated analysis of the Telegram ecosystem both necessary and complementary to existing studies.

Most recently, KeyMagnet~\cite{keymagnet2025} conducted the largest-scale credential leakage study in the app-in-app paradigm, analyzing 413,775 mini-apps across six super-app platforms and detecting 84,491 credential leaks via semantic graph-matching. Their approach targets platform-issued credentials (e.g., \texttt{AppSecret}, \texttt{accessToken}) within ecosystems that enforce documented authentication workflows. Our work is complementary: we target application-level secrets---wallet mnemonics, private keys, and per-app JWTs---in Telegram's unvetted Mini App environment, where no centralized credential API exists.

Real-world incidents further underscore these risks: the Paragon spyware incident~\cite{bbc_paragon}, the Bybit breach~\cite{bbc_bybit}, CVE-2025-24201~\cite{cveorg2025} (breaking webview boundaries), and the StilachiRAT Trojan~\cite{microsoft_stilachirat} all illustrate the severe consequences of sensitive data mishandling. Exploit marketplaces such as Crowdfense\footnote{\url{https://www.crowdfense.com}} highlight ongoing risks from zero-day vulnerabilities.

In summary, while prior work has laid a solid foundation for understanding Telegram's cryptographic and systemic vulnerabilities, there is a notable gap in research addressing the secure management of sensitive client-side data in Telegram Mini Apps. In this paper, we bridge this gap by investigating local storage practices, presenting empirical findings from a diverse set of Mini Apps, and proposing practical countermeasures.

%%---------------------------------------------------------------------
\section{Technology Background}
\label{sect:technology}

Telegram is a cloud-based messaging service relying on a client-server architecture, with core components in C++ and the proprietary MTProto protocol~\cite{telegram-mtproto}. It is available across Android, Linux, Windows, iOS, macOS, and web\footnote{\url{https://telegram.org/}}. One of Telegram's most significant innovations is the introduction of Mini Apps (also known as Telegram Web Apps), which expanded its functionalities to a dynamic ecosystem of web-based applications. The native self-custodial cryptocurrency Wallet has been activated by more than 100 million US users~\cite{TgramWall}.

Mini Apps enable third-party developers to render custom interfaces directly within Telegram through the WebView component---an embedded browser-like technology. Unlike bots, which function through commands, Mini Apps display rich web pages. However, WebView has fewer built-in security mechanisms than full browsers. Despite standardization efforts~\cite{kandula2024webview}, many vulnerabilities have been discovered~\cite{luo2011attacks,chin2014bifocals,rizzo2018babelview,elzawawy2021vulnerabilities,zhang2022identity}.

\begin{figure}
    \centering
    \includegraphics[width=0.9\linewidth]{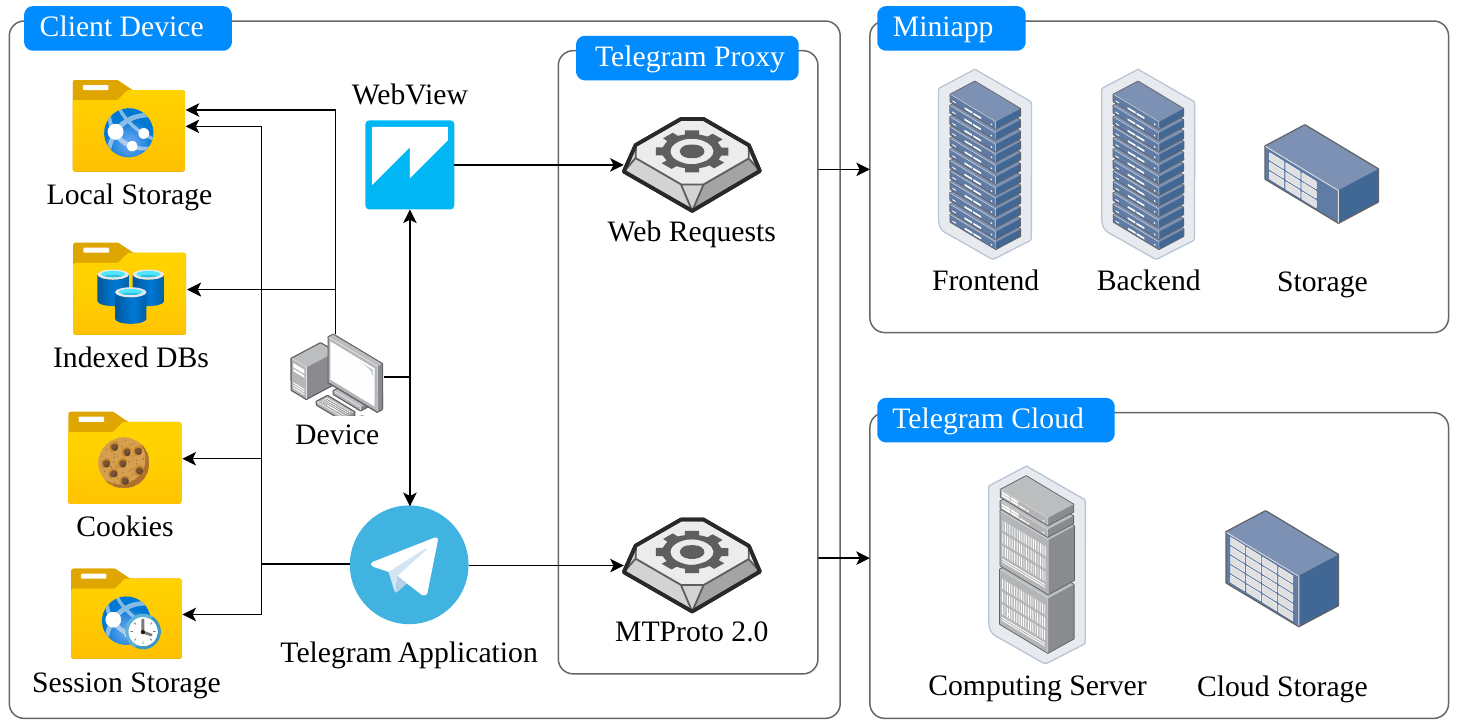}
    \caption{Telegram Mini Apps architecture overview.}
    \Description{A diagram illustrating the dual-channel architecture of Telegram Mini Apps, showing the Telegram client, embedded WebView, Mini App backend, and Telegram cloud infrastructure with their respective communication flows.}
    \label{fig:architecture-overview}
\end{figure}

\subsection{Architecture of Telegram Mini Apps}
As illustrated in~\cref{fig:architecture-overview}, the architecture distinguishes between two parallel communication channels. One channel is dedicated to Telegram-specific communication between the Telegram client and the Telegram cloud infrastructure, secured using the MTProto protocol. The other channel is dedicated to Mini App interactions, rendered within an embedded WebView component. This WebView allows Mini Apps to display web-based content and interact with external backends without leaving the Telegram environment. This dual-flow architecture ensures that Telegram's core security measures remain intact while providing developers the flexibility to implement rich, dynamic interfaces.

\subsection{Data Storage and Integration Flow}
The architecture also highlights the different data storage mechanisms available to Telegram Mini Apps: (i) Cookies ($\sim$4~KB, sent with HTTP requests); (ii) Session storage (ephemeral, cleared on tab close); (iii) IndexedDB (NoSQL, supporting advanced queries); and, (iv) Local Storage (persistent key-value pairs using SQLite~\cite{bhosale2015sqlite} on Linux/macOS or LevelDB~\cite{levelDB} on Windows/Android).

The integration flow begins when a user launches a Mini App within Telegram. The Telegram client initializes an embedded WebView that renders the Mini App's interface. As the user interacts with the Mini App, the front-end sends web requests to the Mini App's backend, while Telegram's APIs concurrently manage user data and session security.

%%---------------------------------------------------------------------
\section{Mini Apps Security Analysis}
\label{sect:vuln_analysis}

The first step in our investigation involved analyzing the standard development process of a Mini App using Telegram's official libraries. Mini Apps follow conventional web development principles and can be built using any web technology stack. The only requirement is using Telegram's SDK to integrate platform-specific functionalities. The official SDK implementation is primarily developed in JavaScript for core functionalities, with TypeScript used for specific graphical components. The SDK is publicly available on GitHub~\cite{twa_github}, as an npm package~\cite{twa_sdk}, or as an injectable script\footnote{\url{https://telegram.org/js/telegram-web-app.js}}.

To investigate potential security threats, we developed a mock Mini App using the Vite framework\footnote{\url{https://docs.ton.org/v3/guidelines/dapps/tma/tutorials/app-examples}}. Our exploration revealed that Telegram imposes no strict conditions on distributing Mini Apps, nor does it enforce security checks such as HTTPS warnings. Additionally, our developed Mini App can inject arbitrary values into local storage, simulating sensitive data such as session tokens. We also implemented a text input field that reflects its content directly on the page without proper sanitization, thereby introducing a reflected Cross-Site Scripting (XSS) vulnerability. \HIGHLIGHT{This setup demonstrates that a remote attacker---without physical device access---could exploit XSS in a vulnerable Mini App to exfiltrate secrets stored under that Mini App's origin (e.g., JWTs or mnemonic phrases) to an external server. This origin-scoped scenario is distinct from the user-space filesystem extraction attack described below.}

Since Telegram does not publish official Mini App usage statistics, we leveraged Trending Apps\footnote{\url{https://tapps.center}} (tApps), a community-driven platform categorizing applications into six distinct sections: ``Games,'' ``Trending,'' ``Web3,'' ``Management,'' ``Utilities,'' and ``Play to Earn''. We collected data for all 683 Mini Apps distributed as shown in \cref{tab:categories_distribution}\footnote{Data updated as of 14th August 2025}. To mitigate selection bias, we applied a stratified, popularity-weighted sampling strategy: we selected the top-ranked Mini Apps by monthly active users within each category, with oversampling of the Web3 and Wallet subcategories due to their higher financial exposure and thus greater adversarial interest. All Mini Apps in our sample had verifiable presence on tApps and a functional desktop client, ensuring comparability of results. Apps were excluded only for objective, pre-defined criteria like inability to load on any tested platform, mobile-only operation, or absence of any persistent storage interaction (verified by running the app and inspecting its storage directory after a standardized interaction sequence).

\begin{table}
    \centering
    \caption{Total unique Mini Apps. $^{\ast}$Not considering Mini Apps belonging to more than one category.}
    \Description{Distribution of 763 (683 unique) Mini Apps across six categories.}
    \label{tab:categories_distribution}
    \resizebox{0.9\linewidth}{!}{%
    \begin{tabular}{l@{\hspace{0.5cm}}||l@{\hspace{0.5cm}}l@{\hspace{0.5cm}}l@{\hspace{0.5cm}}l@{\hspace{0.5cm}}l@{\hspace{0.5cm}}l}
        \hline
        \textbf{Category} & Web3 & Games & Utilities & Play to Earn & Management & Trending \\
        \hline
        \addlinespace[0.1cm]
        \textbf{Elements} & 204 & 200 & 200 & 106 & 41 & 12 \\
        \addlinespace[0.1cm]
        \textbf{Percentage} & 26.74 & 26.21 & 26.21 & 13.89 & 5.37 & 1.57 \\
        \addlinespace[0.1cm]
        \hline
        \addlinespace[0.1cm]
        \textbf{Total} & \multicolumn{6}{c}{\textbf{763 (683$^{\ast}$)}} \\
        \addlinespace[0.1cm]
        \hline
    \end{tabular}
    }
\end{table}

\subsection{Threat Model}
The threat model focuses on the client-side execution environment of Telegram Mini Apps and the mechanisms by which sensitive data is stored on the user's device. Three software components are involved: (i) the \emph{Telegram Client}, which authenticates the user and embeds a WebView; (ii) the \emph{Embedded WebView}, which executes Mini App code written by third-party developers; and, (iii) the \emph{Local Storage}, where the WebView persists key-value data such as session identifiers, wallet information, or application state.

\medskip
\noindent\textbf{Attacker capabilities.}
We consider adversaries who obtain \emph{logical} or \emph{physical} access to the victim's device. These conditions arise in realistic circumstances such as temporary device handover---e.g.\ during charging in public places, short-term access by acquaintances, device theft, or the presence of commodity malware with user-level privileges. Importantly, on the desktop version of Telegram, none of these scenarios require rooting or breaking Telegram's sandboxing; they merely require the ability to read files from the user's profile directory---an action well within reach of standard spyware, infostealers, and widely reported stalkerware families.

Once such access is achieved, the attacker can browse, copy, and exfiltrate the files used by Telegram's WebView for Mini App storage. Although Telegram provides an optional access PIN that protects the user interface, this mechanism \emph{does not encrypt} the underlying storage. Therefore, any Mini App that stores sensitive values in plaintext leaves these artefacts immediately accessible to the attacker.

\medskip
\noindent\textbf{Attack scenarios.}
An attacker can, even with a brief physical access window, copy the Mini App storage directory and extract secrets offline. More critically, modern infostealers---such as RedLine, Raccoon, and Vidar, which collectively compromise millions of devices annually---routinely harvest browser data, LevelDB/SQLite stores, and application cache directories without requiring root privileges~\cite{10.1145/3133956.3134067}. Since Mini App storage uses the same underlying technologies (LevelDB on Windows/Android; SQLite on Linux/macOS), such malware can trivially exfiltrate these database files from the User space. Importantly, unlike traditional browser cookies that may benefit from at-rest encryption in modern browsers (e.g., Chrome's App-Bound Encryption), Telegram's WebView storage offers no such protection---making Mini App data a softer target than equivalent data in standalone browsers~\cite{kandula2024webview}. \HIGHLIGHT{Separately, a web attacker who exploits XSS in a specific Mini App can read and exfiltrate that Mini App's origin-scoped WebView storage; same-origin boundaries do not grant access to storage belonging to unrelated Mini Apps.} Finally, through social engineering attacks users may be tricked into installing software that requests filesystem access or inserting malicious devices. Such applications and devices can extract Mini App databases and send sensitive data to a remote server.

\subsection{Cryptocurrency-related Mini Apps}
Unlike traditional mobile applications, Mini Apps do not require installation---they are web applications accessible via a unique Telegram URL. We selected 40 of the most common Mini Apps across the ``Web3'', ``Games'', and ``Utilities'' categories, with a particular focus on ``Web3'' due to its frequent handling of sensitive financial data. We additionally prioritized cryptocurrency-related Mini Apps linked to rewarding or mining systems, as financial incentives make them a prime target for adversaries.

A preliminary result revealed that any running Mini App inherits the permissions granted to the Telegram app, with no additional restrictions on the WebView environment. Additionally, since Mini Apps use cookies, GDPR compliance~\cite{eprivacyDirective2002} is required but largely absent. \HIGHLIGHT{Among the 61 Mini Apps considered, 37 met the processing criteria and were analyzed. Of these, 30 demonstrated poor handling of Local Storage: 17 exposed session tokens, JWTs, or equivalent replayable authentication artifacts, 16 out of 30 persistently stored private keys or secrets, and 5 out of 30 stored mnemonic recovery phrases---enabling full wallet compromise.}

These vulnerabilities arise because Local Storage is easily accessible from the device. In most cases, data was stored in plaintext; even when encryption was applied, keys and decryption parameters were stored alongside the ciphertext or hardcoded in the Mini App code, rendering the protection ineffective.

Notably, PocketFi---counting 3,457,131 subscribed users\footnote{\label{data_updated}Data updated as of 14th August 2025}---manages all operations through an authentication header leveraging raw Telegram user data rather than conventional session tokens, relying on static plaintext data for authentication~\cite{6234436}. These findings highlight significant security flaws in how Mini Apps handle sensitive user data.

\subsection{Wallet Mini Apps}
Following the comprehensive analysis, we focused on a subset of the Mini Apps most widely used by the Telegram community. In particular, we selected those Mini Apps that are more likely to be targeted by attackers due to their financial pay-off potential. Hence, the emphasis was placed on Mini Apps for executing cryptocurrency transactions, commonly called App Wallets.

Cryptocurrencies are attractive targets for malicious actors, primarily due to their digital and decentralized characteristics. In this context, Wallets represent a particularly enticing target because any security flaw within these systems can potentially be exploited to compromise cryptocurrency assets.

With the introduction of the TON blockchain, Telegram integrated a default ``Wallet'' Mini App functioning as a Web3 wallet aggregator~\cite{TON2025ExclusivePartnership}, alongside {\em TON Connect} for wallet interoperability.

TON Connect supports 21 App Wallets\textsuperscript{\ref{data_updated}}. We excluded 12 that are hyperlinks to external applications and one requiring identity verification, leaving nine fully integrated App Wallets: Bitget Wallet Lite, DeWallet, Fintopio, HOT, OKX Mini Wallet, Tobi, Tomo Wallet, Tonkeeper, and Wallet. TENET identified sensitive data in all except Tonkeeper.

\subsubsection{Bitget Wallet Lite}
This non-custodial wallet serves over 6 million users. TENET extracted the JWT from local storage and it was used to perform reverse engineering. It revealed the mnemonic key can be decrypted using a ``Salt'' from \verb|/userv2/v1/uc/tgwFetchSalt|, where the initialization vector derives from a SHA256 hash of the six-digit password, device UUID, and Telegram ID stored locally. The ``seqnum'' timestamp parameter appears susceptible to replay attacks.

\subsubsection{DeWallet}
TENET extracted the JWT session token. At least one publicly documented API endpoint\footnote{\url{https://battery.dewallet.pro/docs}} (\path{/api/v1/users/balance}) was confirmed operational. Further endpoints such as \path{/api/v1/orders/\{order_id\}/pay} could uncover additional exploitation vectors.

\subsubsection{Fintopio}
This cryptocurrency wallet streamlines the payment process within Telegram and has achieved significant user adoption. TENET located the JWT token within local storage. We then performed a successful escalation of privileges by using the extracted token to query the endpoint \verb|/non-custodial/seed/cloud|. The server responded with the fully decrypted mnemonic key of the Wallet---meaning that an attacker who extracts the JWT from local storage can recover the complete wallet seed without needing any additional credentials, resulting in irreversible financial compromise.

\subsubsection{HOT}
HOT Wallet functions both as a non-custodial wallet and as a token, within the NEAR ecosystem~\cite{near_ecosystem}---a blockchain designed to be fast, secure, scalable, and carbon-neutral. Unlike other cases, TENET detected an encrypted text fragment within the local storage, following the structure \verb|iv:ciphertext|, encoded in hexadecimal format.
By performing reverse engineering on the obfuscated JavaScript code executed by the Mini App within Telegram, we determined that the encryption algorithm used was AES-CBC with SHA512. However, the analysis revealed that the derivation key could be fully recovered, as the Mini App's code explicitly contained both the password and key derivation parameters. The password was hardcoded: \blackout{secret-text}\footnote{\label{redaction_note}Critical information redacted; redaction will be removed when Mini App owners confirm the vulnerability is resolved}, nullifying any security benefits of password-based encryption. The salt value was also trivially predictable: \blackout{salt}\footnoteref{redaction_note}. Finally, the number of PBKDF2 iterations was set to 1---four orders of magnitude below the OWASP-recommended 210,000 for SHA512\footnote{\label{cheatsheet_url}\url{https://cheatsheetseries.owasp.org/cheatsheets/Password_Storage_Cheat_Sheet.html}}. Due to these weaknesses, decrypting the stored data was straightforward, leading to the extraction of both the Wallet's mnemonic key and the JWT token---achieving full wallet compromise.

\subsubsection{OKX Mini Wallet}
Formerly known as OKEx, this major cryptocurrency exchange has developed its own Wallet as a Telegram Mini App (OKX Mini Wallet). During the analysis, a ``vault'' property was discovered in local storage, containing an AES-GCM/SHA256 encrypted value encoded in Base64. A custom RegEx pattern was designed to extract this value and integrated into TENET. The salt value and the iteration count for key derivation were directly extractable by decoding the Base64-encoded ``vault'' property. Through reverse engineering, it was determined that the required password was derived by concatenating the user-defined password with three environment variables: ``Operating System Name'', ``Runtime ID'', and ``CPU Name''. Empirical testing revealed these always resolve to static values: win$~\leftarrow$~Operating System Name, okx-wallet$~\leftarrow$~Runtime\_ID, and ``undefined'' for CPU Name (inaccessible from user space). This predictable pattern significantly reduces the password's entropy, making brute-force attacks feasible. The iteration count was consistently below 60,000---only one-tenth of the OWASP-recommended 600,000 for PBKDF2-HMAC-SHA256\footnoteref{cheatsheet_url}. Due to these weaknesses, it was possible to decrypt the stored data, brute-force the password, and obtain the Wallet's mnemonic---again, achieving full wallet compromise.

\subsubsection{Tobi, Tomo Wallet, and Tonkeeper}
In the analysis of Tobi (a wallet integrating an AI assistant), our tool extracted the JWT token and verified it via \verb|/v2/wallet/portfolio/balance|. For Tomo Wallet (an all-in-one crypto social wallet under active development), TENET identified the JWT token for internal service communications, verified via \verb|/tg-auth/v1/token/balance|. Tonkeeper, a non-custodial wallet notable for its open-source nature\footnote{\url{https://github.com/tonkeeper}}, was the only case where our analysis did not identify any concrete exploitable sensitive information.

\subsubsection{Wallet}
This application is the officially supported Wallet integrated within Telegram---it appears as an entry in Telegram's side menu. Despite being the Wallet with the most extensive user base (accessible to all Telegram users), it exhibits one of the most critical vulnerabilities. The mnemonic key, which can be used to reset the account by bypassing password authentication completely, is stored in plaintext within local storage. TENET detected the presence of the mnemonic key, thereby exposing a critical security flaw.

\subsection{Remote Data Exfiltration PoC}
\HIGHLIGHT{As a distinct user-space attack vector, drawing on CYFIRMA's RenderShock zero-click technique~\cite{rendershock}, we embedded a Visual Basic macro in a fake cryptocurrency-management spreadsheet\footnote{Available in the TENET repository} that silently compresses the user's \texttt{tdata} directory and uploads it to an attacker-controlled endpoint.} Since Mini App data resides in a user-accessible folder and Excel macros execute with the logged-in user's privileges, no elevation is required. This proof-of-concept shows that a single office file is enough to extract complete Telegram wallet data.

%%---------------------------------------------------------------------
\section{TENET}
\label{sect:tool_architecture}

In this section we detail the rationale followed by TENET, as well as its architecture, and execution flow.

\subsection{TENET Rationale and Architecture}
TENET is a command-line tool designed to identify sensitive data stored insecurely within the local storage of devices running Telegram. It autonomously detects the device type, operating system, and the presence of multiple accounts, then locates Mini App storage directories and identifies the database type (LevelDB or SQLite).

\begin{figure}
    \centering
    \includegraphics[width=0.8\linewidth]{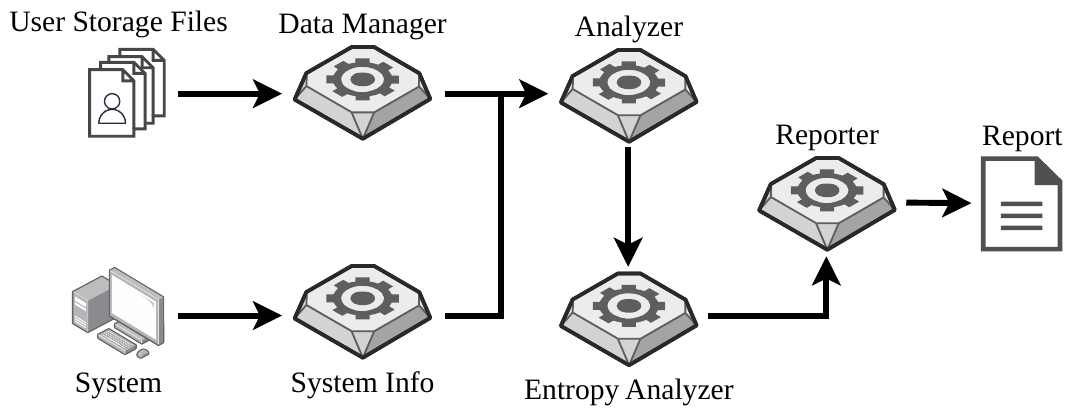}
    \caption{Component diagram of TENET.}
    \Description{A component diagram showing TENET's modular architecture, with an Analyzer orchestrating the SystemInfo, DataManager, EntropyAnalyzer, and Reporter components.}
    \label{fig:tool-architecture}
\end{figure}

TENET extracts all relevant data and applies over 27 custom RegEx patterns to identify sensitive elements. Two refinement mechanisms minimize false positives: (i) Charset Validation against predefined character sets; and, (ii) Entropy-Based Filtering using customizable randomness thresholds detailed in \cref{tab:charset_randomness_patterns}. The tool generates reports in plaintext, CSV, or JSON format, and supports configurable parameters for customization. Its modular architecture (\cref{fig:tool-architecture}) comprises independent components orchestrated by a central Analyzer.

\subsubsection{Design Rationale for Patterns, Thresholds, and Filters}
Each of TENET's 27 detection patterns was derived from one of three sources: (i) established credential formats documented in prior secret-detection literature~\cite{keymagnet2025,steffens2019dont} (e.g., JWT three-part Base64URL structure, AWS AKIA prefix); (ii) empirical analysis of Mini App JavaScript source code and storage dumps collected during a pre-study phase; or (iii) format specifications from the relevant protocol or service documentation (e.g., BIP-39 for mnemonic phrases, RFC 7519 for JWTs).

\noindent\textbf{Entropy thresholds.}
Shannon entropy is used to distinguish genuine high-entropy secrets from low-entropy strings that syntactically resemble them. Thresholds were set per pattern class based on the observed entropy distribution of true positives versus false positives in a held-out validation set of 200 manually labeled storage entries (100 genuine secrets, 100 benign strings). High-entropy classes (JWTs, session tokens, private keys, OAuth tokens) use a threshold of 0.65 bits/symbol normalized to [0,1], which rejected 91\% of false positives while retaining 98\% of true positives in validation. Low-entropy classes (mnemonics, pin codes, S3 bucket names) use lower thresholds (0.0--0.2) reflecting their structured, human-readable nature; for these, the RegEx structural constraint carries the primary discriminative load.

\noindent\textbf{Charset validation.}
Charset validation is applied as a complementary filter to catch structurally matching strings whose character distribution is inconsistent with the expected encoding. For example, a BASE64URL-encoded JWT payload must use only \texttt{[A-Za-z0-9\textunderscore-]}; any match containing characters outside this set is discarded. This eliminates a significant class of false positives arising from storage keys that happen to contain three dot-separated substrings. Charset sets were defined based on the encoding specifications of each secret type.

\noindent\textbf{Pattern coverage.}
The 27 patterns cover six functional categories: (1)~\emph{Authentication tokens} (JWT, OAuth, session tokens, generic tokens); (2)~\emph{Cryptographic material} (private keys, mnemonic phrases, encrypted data, encryption parameters, vault data); (3)~\emph{Cloud and API credentials} (AWS access keys, AWS ARNs, S3 buckets, Google API keys, GitHub tokens); (4)~\emph{Database credentials} (connection URIs, username/password pairs); (5)~\emph{PII} (email addresses, IP addresses, Telegram IDs, PIN codes); and (6)~\emph{Application secrets} (URL-embedded credentials, WebSocket URLs, sensitive parameters, encrypted references). This taxonomy ensures coverage of the full spectrum of sensitive artifacts that prior work has identified in web and mobile application storage~\cite{hanna2010emperor,steffens2019dont,keymagnet2025}.

\begin{table}[h]
    \centering
    \caption{Pattern charset and randomness details for optimal data match.}
    \Description{Table listing 27 TENET pattern types with their associated character sets and randomness thresholds.}
    \label{tab:charset_randomness_patterns}
    \footnotesize
    \resizebox{0.7\linewidth}{!}{%
    \begin{tabular}{l@{\hspace{\tabcolsep}\hspace{0.3cm}}l@{\hspace{\tabcolsep}\hspace{0.3cm}}l}
        \toprule
        \textbf{Name} & \textbf{Charsets} & \textbf{Randomness}\\
        \midrule
        JWT & BASE64URL, PRINTABLE & 0.65 \\
        \hline
        OAUTH\_TOKEN & BASE64URL, ALPHANUMERIC & 0.65 \\
        \hline
        SESSION\_TOKEN & BASE64, HEX, ALPHANUMERIC & 0.65 \\
        \hline
        MNEMONIC & PRINTABLE & 0.1 \\
        \hline
        MNEMONIC\_REFERENCE & PRINTABLE & 0.1 \\
        \hline
        ENCRYPT\_REFERENCE & BASE64, PRINTABLE & 0.65 \\
        \hline
        PRIVATE\_KEY & BASE64, PRINTABLE & 0.65 \\
        \hline
        ENCRYPTED & BASE64, PRINTABLE & 0.65 \\
        \hline
        AWS\_ACCESS\_KEY & UPPERCASE & 0.65 \\
        \hline
        GOOGLE\_API\_KEY & ALPHANUMERIC & 0.65 \\
        \hline
        GITHUB\_TOKEN & ALPHANUMERIC & 0.65 \\
        \hline
        DB\_CONNECTION\_URI & PRINTABLE & 0.2 \\
        \hline
        DB\_CREDENTIALS & ALPHANUMERIC, SPECIAL & 0.2 \\
        \hline
        ENCRYPTION\_PARAM & BASE64, HEX & 0.65 \\
        \hline
        URL\_CREDENTIALS & PRINTABLE & 0.65 \\
        \hline
        CREDIT\_CARD & NUMERIC & 0.5 \\
        \hline
        PIN\_CODE & NUMERIC & 0.0 \\
        \hline
        S3\_BUCKET & LOWERCASE & 0.0 \\
        \hline
        AWS\_ARN & PRINTABLE & 0.0 \\
        \hline
        GENERIC\_TOKEN & ALPHANUMERIC & 0.3 \\
        \hline
        SENSITIVE\_PARAM & ALPHANUMERIC, SPECIAL, PRINTABLE & 0.2 \\
        \hline
        IP\_ADDRESS & NUMERIC & 0.2 \\
        \hline
        EMAIL\_ADDRESS & PRINTABLE & 0.2 \\
        \hline
        WEB\_SOCKET\_URL & PRINTABLE & 0.2 \\
        \hline
        ENCRYPTED\_DATA & BASE64, HEX, PRINTABLE & 0.2 \\
        \hline
        VAULT & ALPHANUMERIC, SPECIAL, PRINTABLE & 0.0 \\
        \hline
        TGID & PRINTABLE & 0.0 \\
        \bottomrule
    \end{tabular}
    }
\end{table}

After the analysis, the tool generates a structured report that can be used for supporting continuous security assessments by allowing integration into CI/CD pipelines and existing security test suites.

\subsection{TENET Execution Flow}
\begin{figure}
    \centering
    \includegraphics[width=\linewidth]{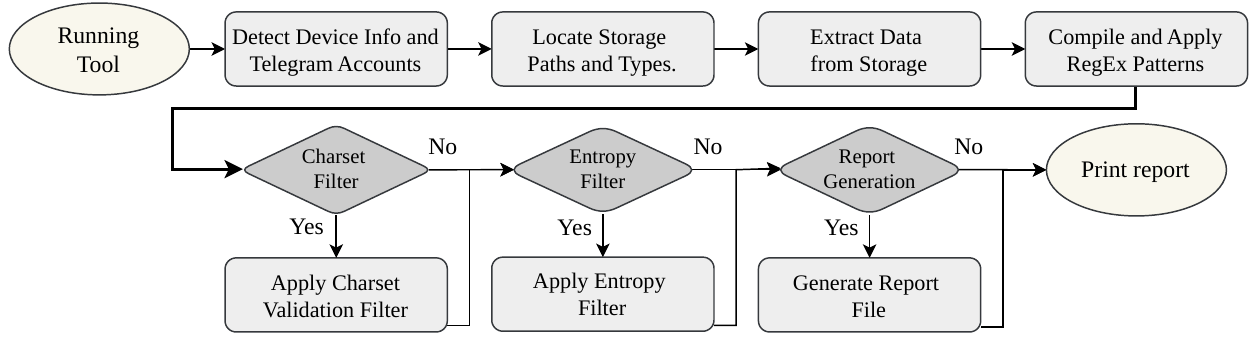}
    \caption{TENET's flow diagram.}
    \Description{A flowchart illustrating TENET's execution pipeline: OS detection, database identification, data extraction, RegEx matching, entropy filtering, and structured report generation.}
    \label{fig:flow-chart}
\end{figure}

The execution flow (\cref{fig:flow-chart}) proceeds as follows: the Analyzer uses SystemInfo to detect the OS and databases, then DataManager extracts local storage data from all accounts. Patterns are matched using pre-compiled RegEx, refined by the EntropyAnalyzer (evaluating randomness and charset compliance), and output via the Reporter.

A simplified Python PoC script was also developed using only built-in libraries. Both versions operate in real-time with minimal resources, requiring only reading access to Telegram's data. The full version requires \verb|pydantic|; \verb|pyjwt| is optional for JWT decoding.

\subsection{TENET Evaluation}
\label{sect:tenet_eval}

To assess TENET's detection quality, we constructed a ground-truth evaluation dataset independently of the main measurement study. We manually annotated 500 storage entries extracted from 20 Mini Apps not included in our main dataset: 243 were labeled as true sensitive artifacts by two independent annotators (inter-rater agreement $\kappa = 0.91$), and 257 as benign strings. TENET was then run on this dataset in fully automated mode.

\noindent\textbf{Detection performance.}
\HIGHLIGHT{TENET achieved a \emph{precision} of 94.3\%, a \emph{recall} of 96.7\%, and an F1-score of 95.5\%. The 5.7\% complement of precision is the false-discovery rate, whereas the 3.3\% complement of recall is the false-negative rate. The false positives consisted primarily of generic alphanumeric storage keys whose length and entropy fell within the threshold band for GENERIC\_TOKEN; the false negatives included two obfuscated mnemonic phrases encoded as comma-separated integers rather than BIP-39 words. Both failure modes are documented as known limitations. Because prior tools were evaluated on different corpora and with different error definitions, their reported error rates are not treated as a direct quantitative baseline for TENET.}

\noindent\textbf{Runtime and scalability.}
TENET completes a full single-account scan (including database extraction, pattern matching, entropy analysis, and report generation) in approximately $2.1 \pm 0.3$ seconds averaged over 30 runs on the AMD Ryzen 9 5900X testbed, with linear scaling observed up to 50 simultaneously installed Mini Apps. Multi-account mode adds approximately 1.8 seconds per additional account.

\noindent\textbf{Comparison with existing tools.}
General-purpose secret scanners (TruffleHog~\cite{keymagnet2025}, detect-secrets) are designed for source code repositories and cannot parse LevelDB or SQLite databases. File system credential harvesters used in infostealer malware target browser-specific paths and formats, not Telegram's WebView storage layout. TENET fills this gap by combining Telegram-specific storage path resolution, multi-database backend support (LevelDB/SQLite), and a pattern library calibrated to the secret types prevalent in Mini App ecosystems.

\begin{table}[h]
    \centering
    \caption{Details of how many cryptocurrency-related and Wallet Mini Apps we considered for our analysis and how many were excluded.}
    \Description{Breakdown of Mini Apps considered and excluded across two categories.}
    \label{tab:analyzed_mini_apps}
    \resizebox{0.6\linewidth}{!}{%
    \begin{tabular}{l@{\hspace{1cm}}c@{\hspace{0.8cm}}c}
        \toprule
         & Cryptocurrency-related & Wallet \\
        \midrule
        Non-processable & 3 & 1 \\
        \hline
        Mobile Only & 3 & 0 \\
        \hline
        No functionalities & 6 & 11 \\
        \hline
        Analyzed & 28 & 9 \\
        \hline
        \textbf{Total} & 40 & 21 \\
        \bottomrule
    \end{tabular}
    }
\end{table}

%%---------------------------------------------------------------------
\section{Experimental Results}
\label{sect:statistics}

This section describes the TENET experimental setup, outlines the criteria for selecting Telegram Mini Apps, the metrics for assessment, and the overall methodology employed in our experiments. Finally, we report our findings and show possible fixes to the highlighted vulnerabilities.

\subsection{Testbed}
\label{sect:testbed}
We executed all analyzed Mini Apps on a temporary Telegram account using Desktop clients (Windows 5.12.3, macOS 5.13.1, Linux 5.13.1) and Android mobile client (v11.13.2). Tests ran on an AMD Ryzen 9 5900X with 64GiB RAM (Windows/Linux dual boot), an Apple Mac mini M1 (macOS), and an Android API 36 VM with root privileges. We subsequently executed the same tests on a second Telegram account to verify TENET's multi-account detection capability.

\subsection{Results Analysis}
\label{sect:results}

As \cref{tab:analyzed_mini_apps} shows, we filtered out Mini Apps that did not execute correctly, are mobile-only, or lack functionality. Of 40 cryptocurrency-related Mini Apps initially considered, 12 were excluded (three failed to load, three were mobile-only, six lacked sensitive data functionality), leaving 28 for analysis with TENET.

\begin{table}
    \centering
    \caption{Overall details of what type of data we found from the analyzed cryptocurrency-related and Wallet Mini Apps.}
    \Description{Summary of data types found across analyzed Mini Apps.}
    \label{tab:found_in_mini_apps}
    \resizebox{0.60\linewidth}{!}{%
    \begin{tabular}{l@{\hspace{1cm}}c@{\hspace{0.8cm}}c}
        \toprule
         & Cryptocurrency-related & Wallet \\
        \midrule
        Nothing found & 7 & 1 \\
        \hline
        Session Tokens/JWT & 11 & 6 \\
        \hline
        Mnemonic Phrase & 1 & 4 \\
        \hline
        Encrypted Data & 2 & 2 \\
        \hline
        Sensitive params & 13 & 3 \\
        \bottomrule
    \end{tabular}
    }
\end{table}

Moreover, a total of 21 Wallet Mini Apps were considered. We were able to execute our tool and further investigate the security on a subset of 9 Wallets only. Indeed, out of the originally considered 21 Wallet Mini Apps, one was not processable because it required an official user identification proof, and 11 of them were simply redirection URLs to the official external Wallet.

\begin{table*}[!h]
    \centering
    \caption{Finding of our analysis after running TENET. $^{\ast}$Required manual analysis using the session token through data decryption and API invocation. $^{\dagger}$Mini Apps not showing data flows.}
    \Description{Detailed per-app findings table showing which sensitive data types TENET detected in each analyzed Mini App.}
    \label{tab:mini_app_findings}
    \resizebox{\textwidth}{!}{%
    \setlength{\extrarowheight}{0.2em}
    \begin{tabular}{c l c c c c c c c c}
        \toprule
        & \textbf{Title}
        & \makecell{\textbf{JWT}}
        & \makecell{\textbf{Mnemonic}\\\textbf{phrase}}
        & \makecell{\textbf{Sensitive}\\\textbf{parameters}}
        & \makecell{\textbf{Email}\\\textbf{address}}
        & \makecell{\textbf{Encrypted}\\\textbf{data}}
        & \textbf{Vault}
        & \textbf{Tgid}
        & \makecell{\textbf{Encrypted}\\\textbf{reference}} \\
        \midrule
        \multirow{9}{*}{\rotatebox{90}{\textbf{Wallet Mini Apps}}}
        & Bitget Wallet Lite   & \boldcheckmark &  & \boldcheckmark & \boldcheckmark &  &  &  & \\ \cline{2-10}
        & DeWallet             & \boldcheckmark &  &  &  &  &  &  & \\ \cline{2-10}
        & Fintopio             & \boldcheckmark & \boldcheckmark$^{\ast}$ &  &  &  &  &  & \\ \cline{2-10}
        & HOT Wallet           &  & \boldcheckmark$^{\ast}$ &  &  & \boldcheckmark &  &  & \\ \cline{2-10}
        & OKX Wallet           & \boldcheckmark & \boldcheckmark$^{\ast}$ &  &  & \boldcheckmark & \boldcheckmark &  & \\ \cline{2-10}
        & Tobi                 & \boldcheckmark &  & \boldcheckmark &  &  &  &  & \\ \cline{2-10}
        & Tomo Wallet          & \boldcheckmark &  & \boldcheckmark & \boldcheckmark &  &  & \boldcheckmark & \\ \cline{2-10}
        & Ton Keeper$^{\dagger}$ &  &  &  &  &  &  &  & \\ \cline{2-10}
        & Wallet               &  & \boldcheckmark &  & \boldcheckmark &  &  &  & \boldcheckmark \\
        \midrule
        \multirow{21}{*}{\rotatebox{90}{\textbf{Mini Apps}}}
        & ArbitrageScanner Wallet  & \boldcheckmark &  & \boldcheckmark &  &  &  &  & \\ \cline{2-10}
        & ArchiteTon               &  &  &  &  &  &  &  & \boldcheckmark \\ \cline{2-10}
        & Bivreost Wallet          &  & \boldcheckmark &  &  & \boldcheckmark &  &  & \\ \cline{2-10}
        & Blum                     &  &  & \boldcheckmark &  &  &  &  & \\ \cline{2-10}
        & Catizen                  & \boldcheckmark &  & \boldcheckmark &  &  &  &  & \\ \cline{2-10}
        & Community                & \boldcheckmark &  &  &  &  &  &  & \\ \cline{2-10}
        & CryptoRank               &  &  & \boldcheckmark &  &  &  &  & \\ \cline{2-10}
        & Fanton Fantasy Football  & \boldcheckmark &  & \boldcheckmark &  &  &  &  & \\ \cline{2-10}
        & Gate.io                  &  &  & \boldcheckmark &  &  &  &  & \\ \cline{2-10}
        & HunteX                   & \boldcheckmark &  &  &  &  &  &  & \\ \cline{2-10}
        & MuggleLink               & \boldcheckmark &  &  &  &  &  &  & \\ \cline{2-10}
        & Notcoin                  & \boldcheckmark &  &  &  &  &  &  & \\ \cline{2-10}
        & Pixel Wallet             &  &  & \boldcheckmark &  &  &  &  & \\ \cline{2-10}
        & Playnation               & \boldcheckmark &  & \boldcheckmark &  &  &  &  & \\ \cline{2-10}
        & ROoLZ                    & \boldcheckmark &  & \boldcheckmark &  &  &  &  & \\ \cline{2-10}
        & Telenova: Polkadot Wallet &  &  &  &  &  &  &  & \\ \cline{2-10}
        & The Pixels               &  &  & \boldcheckmark &  &  &  &  & \\ \cline{2-10}
        & Tradoor Trading          &  &  & \boldcheckmark &  &  &  &  & \\ \cline{2-10}
        & Wave Wallet              &  &  &  &  & \boldcheckmark &  &  & \\ \cline{2-10}
        & Yescoin                  & \boldcheckmark &  & \boldcheckmark &  &  &  &  & \\ \cline{2-10}
        & Yescoin \O               &  &  & \boldcheckmark &  &  &  &  & \\
        \bottomrule
    \end{tabular}
    }
\end{table*}

The 40 cryptocurrency-related samples were selected based on popularity and categorical relevance, focusing on the Web3 category. The 21 Wallets were identified from Telegram's recommended wallet list\footnote{\url{https://tapps.center}}.

\HIGHLIGHT{As summarized in \cref{tab:found_in_mini_apps}, 7 of 28 cryptocurrency-related Mini Apps and 1 of 9 Wallets did not expose sensitive data. The remaining 30 of the 37 analyzed Mini Apps (81.1\%; 49.2\% of the 61 initially considered) showed the presence of locally stored data that posed security concerns. After manually filtering TENET false positives, our results (\cref{tab:mini_app_findings}) reveal a systematic pattern of insecure storage: 57\% of the affected Mini Apps (17 out of 30) expose session tokens, JWTs, or equivalent replayable authentication artifacts; 17\% (5 out of 30) save mnemonic phrases without robust encryption, with 4 out of 9 wallet applications directly exposing the wallet seed; and 20\% (6 out of 30) employ encryption whose keys, salts, and iteration parameters are stored alongside the ciphertext, nullifying any cryptographic protection. The explicit JWT instances reported in \cref{tab:mini_app_findings} were verified to be valid and replayable against the respective backend APIs (e.g., retrieving account balances or querying wallet seeds). Only \emph{Ton Keeper} revealed no data flows.}

To enable severity-aware analysis, we classify the \textbf{30} vulnerable Mini Apps into three tiers based on the immediate exploitability of the exposed data: (i) Tier 1 --- Plaintext storage (e.g., \textit{Wallet}, \textit{Bivreost Wallet}): mnemonic phrases or private keys stored with no protection whatsoever---a single file copy results in full, irreversible wallet compromise; (ii)Tier 2 --- Recoverable encryption (e.g., \textit{HOT Wallet}, \textit{OKX Wallet}): data is encrypted, but the decryption key, salt, and/or iteration count are stored alongside the ciphertext or hardcoded in the app's JavaScript, making offline decryption trivial; (iii)Tier 3 --- Replayable tokens: JWTs or session tokens stored in plaintext, where exploitation requires a live network connection to replay the token against the backend API, yet causes session hijacking, identity impersonation, and in some cases (e.g., \textit{Fintopio}) full seed recovery via a privileged API endpoint.
Multiple sensitive elements often coexist: \emph{Bitget Wallet Lite} stores a JWT, session token, mnemonic phrase, and email in plaintext; \emph{OKX Wallet} uses encryption and a vault yet remains vulnerable. In \emph{Tomo Wallet}, the user's \emph{Tgid} value---directly linked to the phone number---can be recovered.

\HIGHLIGHT{The affected applications account for an aggregate of more than 53 million monthly active-user counts (cfr.~\cref{tab:monthly_active_users}), with over 85\% associated with Wallet applications such as \textit{Wallet}, \textit{Bitget Wallet Lite}, \textit{HOT Wallet}, and \textit{Tomo Wallet}. Because the same person may use multiple Mini Apps and cross-app overlap statistics are unavailable, this aggregate must not be interpreted as the number of unique affected individuals. Assigning each application to its highest applicable severity tier, approximately 40.4M aggregate monthly active-user counts are associated with Tier-1/Tier-2 vulnerabilities enabling immediate, irreversible asset loss, while approximately 12.6M are associated with Tier-3 session-level risks requiring active network exploitation.}
These figures highlight the accelerating growth of the Mini App ecosystem and the pressing need for robust security measures. Given that many of these apps facilitate financial transactions, the ramifications of improper data handling could be especially severe: assets held in these wallets are exposed to an adversarial DeFi landscape in which financially motivated fraud is systematic and well-documented---rug pull schemes alone have been shown to operate profitably across all major blockchains~\cite{kalal2025rugpull}, meaning that an attacker with access to a plaintext mnemonic phrase faces a target-rich environment of liquid, largely unrecoverable funds.

\begin{table*}[!h]
    \centering
    \caption{Monthly active users for each Mini App. $^{\ast}$Data until 23 June 2026.}
    \Description{Full list of all analyzed Mini Apps with their Telegram URLs, categories, and monthly active user counts.}
    \label{tab:monthly_active_users}
    \resizebox{0.95\textwidth}{!}{%
    \begin{tabular}{lllr}
        \toprule
        \textbf{Title} & \textbf{Telegram URL} & \textbf{Categories} & \textbf{Monthly Active Users}\\
        \midrule
        ArbitrageScanner Wallet & https://t.me/wallet\_arbitragescanner\_bot/ & Web3 & --- \\
        \hline
        ArchitecTon & https://t.me/architec\_ton\_bot & Web3 & --- \\
        \hline
        Battles & https://t.me/battlescryptobot & Games, Trending, Web3 & 1,421 \\
        \hline
        Bit.Store Card & https://t.me/BitStore\_Card\_bot & Web3 & --- \\
        \hline
        Bitget Wallet Lite & https://t.me/BitgetWallet\_TGBot & Web3 & 27,167 \\
        \hline
        BitgetWeb3 & https://t.me/BitgetOfficialBot & Web3 & 26,804 \\
        \hline
        Bivreost Wallet & https://t.me/bivreost\_bot & Web3 & --- \\
        \hline
        Blum & https://t.me/BlumCryptoBot & Games, Web3, Play to Earn & \REV{442,634} \\
        \hline
        Catizen & https://t.me/catizenbot & Games, Web3 & --- \\
        \hline
        Clayton & https://t.me/claytoncoinbot & Web3, Play to Earn & --- \\
        \hline
        Community & https://t.me/community\_bot/ & Web3, Management & 13,772 \\
        \hline
        CryptoRank & https://t.me/CryptoRank\_app\_bot & Trending, Web3 & 42,528 \\
        \hline
        DeWallet & https://t.me/dewallet & Web3 & --- \\
        \hline
        DUCK MY DUCK & https://t.me/duckmyduck\_bot & Games, Web3, Play to Earn & \REV{1,031,898} \\
        \hline
        Fanton Fantasy Football & https://t.me/fantongamebot & Games, Web3 & --- \\
        \hline
        Fintopio & https://t.me/fintopio & Web3 & 7,407 \\
        \hline
        GAMEE & https://t.me/gamee & Games, Web3 & \REV{409,577} \\
        \hline
        Gate.io & https://t.me/gate\_official\_bot & Web3 & --- \\
        \hline
        Gate.io Wallet & https://t.me/gateio\_Wallet\_bot & Web3 & --- \\
        \hline
        Gatto & https://t.me/gatto\_gamebot & Games, Web3 & 7,080 \\
        \hline
        GOSTREAM & https://t.me/biztycoon\_bot & Web3 & --- \\
        \hline
        HOT Wallet & https://t.me/herewalletbot & Web3 & \REV{216,661} \\
        \hline
        HunteX & https://t.me/HuntexBot & Web3 & --- \\
        \hline
        MuggleLink & https://t.me/MuggleLinkBot & Utilities, Play to Earn & --- \\
        \hline
        Notcoin & https://t.me/notcoin\_bot & Games, Web3 & \REV{510,896} \\
        \hline
        Obix & https://t.me/obix\_bot & Web3 & 24,115 \\
        \hline
        OKX Wallet & https://t.me/OKX\_WALLET\_BOT & Web3, Utilities & \REV{---$^{\dagger}$} \\
        \hline
        Pacbot & https://t.me/teampacbot/ & Web3 & --- \\
        \hline
        Pixel Wallet & https://t.me/pixel\_wallet\_bot & Web3 & 39,312 \\
        \hline
        Playnation & https://t.me/Playnation\_bot & Games, Web3 & --- \\
        \hline
        PocketFi & https://t.me/pocketfi\_bot & Web3, Play to Earn & 35,686 \\
        \hline
        Punk City & https://t.me/PunkCity2094bot & Web3 & 10,535 \\
        \hline
        ROoLZ & https://t.me/ROoLZQuest\_bot & Web3, Play to Earn & 33,442 \\
        \hline
        Steamify & https://t.me/steamify\_bot & Utilities, Play to Earn & 35,568 \\
        \hline
        telegram.connect & https://t.me/townwifibot & Web3, Utilities & --- \\
        \hline
        Telenova: Polkadot Wallet & https://t.me/telenova\_app\_bot/ & Web3 & --- \\
        \hline
        The Open League Pools & https://t.me/Open\_league\_bot & Web3 & --- \\
        \hline
        The Pixels & https://t.me/the\_pixels\_bot & Games, Web3 & --- \\
        \hline
        Tobi & https://t.me/TobiCopilotBot & Web3 & --- \\
        \hline
        Tomo Wallet & https://t.me/tomowalletbot & Web3 & 881,124 \\
        \hline
        TON Bridge & https://t.me/TONBridge\_robot & Web3 & --- \\
        \hline
        Tonkeeper & https://t.me/tonkeeper & Web3 & \REV{45,231} \\
        \hline
        TonTopUp & https://t.me/TonTopUp\_bot & Trending, Web3, Utilities & 79,842 \\
        \hline
        Tradoor Trading & https://t.me/tradoor\_io\_bot & Web3 & 79,291 \\
        \hline
        Unknown coin & https://t.me/coin\_unk\_bot & Games, Web3 & 2,525 \\
        \hline
        UXUY Wallet & https://t.me/UXUYbot & Trending, Web3 & 31,219 \\
        \hline
        Wallet$^{\ast}$ & https://t.me/wallet & Utilities & 39,453,322 \\
        \hline
        Wave Wallet & https://t.me/waveonsuibot/ & Web3 & 877,650 \\
        \hline
        Yescoin & https://t.me/theYescoin\_bot & Play to Earn & 95,819 \\
        \hline
        Yescoin \O & https://t.me/realyescoinbot & Web3, Play to Earn & --- \\
        \bottomrule
    \end{tabular}
    }
\end{table*}

%%---------------------------------------------------------------------
\section{Discussion}
\label{sect:discussions_limitation_future}

\textbf{Solutions.}
\HIGHLIGHT{The insecure storage practices described above are concerning because they expose data to a wide range of cybersecurity threats, ranging from victim identification and session hijacking to worst-case scenarios where an attacker can fully recover and access a user's Wallet, thereby posing a severe risk of financial exploitation---an issue compounded by the fact that Wallet Mini Apps account for approximately 85\% of the aggregate 53 million monthly active-user counts associated with the affected applications.}
The core issue stems from the inherent limitations of local storage mechanisms that do not enforce robust security controls. This vulnerability could be mitigated by adopting data encryption techniques at multiple levels. For instance, the Telegram client could extend its lock pin mechanism or implement a server-managed password to protect locally cached data. At the same time, Mini App developers could adopt secure encryption routines that ensure decryption keys are never stored locally. Alternatively, when persistent storage is unnecessary, HTTPOnly cookies (as suggested by Zheng et al.~\cite{zheng2015cookies}) or session storage can provide a more secure solution by automatically clearing data upon session termination.

\noindent\textbf{Impact of our findings.}
\HIGHLIGHT{Our analysis reveals that 30 of the 37 analyzed Mini Apps, drawn from 61 applications initially considered, contain critical security flaws. The affected applications account for more than 53 million aggregate monthly active-user counts, over 85\% of which are associated with Wallet applications; this sum may count the same individual more than once. The discovered vulnerabilities enable attacks with immediate real-world consequences, ranging from JWT replay to the exposure of plaintext mnemonic phrases that grant full, irrevocable control over cryptocurrency wallets. Critically, even Telegram's official Wallet---activated by more than 100 million users---was found to expose wallet recovery mnemonics in cleartext. Since blockchain transactions are irreversible, a single compromise of a mnemonic phrase results in permanent financial loss with no possibility of recovery or chargeback.}

\noindent\textbf{Responsible disclosure.}
We responsibly disclosed the identified vulnerabilities to Telegram and to all affected Mini App developers, providing a detailed report with a working proof-of-concept. Telegram received the full platform-level analysis, while individual developers were notified of application-specific exposures. Sufficient lead time was granted before any public disclosure.

Telegram acknowledged our findings and they introduced two new storage APIs in the Mini Apps platform~\cite{telegram_securestorage}: SecureStorage, which leverages OS-native credential management to encrypt sensitive data, and DeviceStorage, a structured persistent storage mechanism within the Telegram client. The SecureStorage API is explicitly designed for tokens, secrets, and authentication state, ensuring that stored values are encrypted and inaccessible to unauthorized applications---directly addressing the core vulnerability we reported.

\HIGHLIGHT{To verify the platform-level remediation, we re-executed TENET against Telegram's official Wallet. After the patch introduced by Telegram, its recovery mnemonic, formerly recoverable in plaintext, was stored in an encrypted, non-recoverable format and was no longer exposed through local storage extraction. We do not generalize this verification to all third-party Mini Apps, whose remediation depends on developer adoption of the new APIs.}

\noindent\textbf{Limitations.}
Our primary findings target the \emph{Desktop} environment (Windows, macOS, Linux), where Mini App storage resides in user-accessible directories without OS-level encryption. On mobile platforms, Android and iOS sandboxing prevents access without root/jailbreak privileges; however, we verified that on rooted devices the same data can be extracted. Crucially, Mini App code stores data insecurely regardless of platform---the protection, where present, is provided by the OS sandbox, not by the application itself. \HIGHLIGHT{Finally, the filesystem-extraction scenarios require physical or logical user-space access, or execution of the malicious document, but no root privileges on Desktop. The XSS scenario instead requires a vulnerability in the target Mini App and is limited to that application's origin-scoped storage.}

%%---------------------------------------------------------------------
\section{Conclusion}
\label{sect:conclusion}

In this paper, we examined the security of Telegram Mini Apps.
Blending web and native technologies in these applications has led to critical vulnerabilities, notably the plaintext storage of authentication artifacts such as JSON Web Tokens (JWTs), session tokens, and wallet mnemonic phrases. Our investigation revealed that a significant proportion of Mini Apps---especially the Wallet Mini Apps---lack proper encryption measures, thereby exposing millions of users to severe risks, including unauthorized access, user impersonation, and financial losses.

To investigate the breadth and depth of these issues, we developed TENET, a dedicated tool that extracts and analyzes local storage data to identify insecure patterns and quantify vulnerabilities.
\HIGHLIGHT{Our experimental results show that, among 61 Mini Apps initially considered based on popularity and category (prioritizing financial Mini Apps such as cryptocurrency wallets), 37 met the processing criteria and were analyzed, and 30 of these contained critical security flaws. We classify these flaws into three severity tiers: plaintext storage (Tier~1), recoverable encryption (Tier~2), and replayable tokens (Tier~3). Across the 30 affected applications, we identified session tokens or equivalent replayable authentication artifacts in 17 cases, private keys or secrets in 16, and wallet recovery mnemonic phrases in 5.}
To illustrate the severity of the problem, even Telegram's own Wallet suffers from a serious vulnerability that could result in full account compromise. Given Telegram's growing ecosystem and the significant volume of sensitive data processed, our findings underscore the urgent need for a more secure approach to Mini App development---a need validated by Telegram's subsequent adoption of two new secure storage APIs~\cite{telegram_securestorage}.

\section*{Ethics and Privacy Statement}
This research involved passive analysis of data stored by publicly available Telegram Mini Apps on the researchers' own devices, using accounts created exclusively for this study. No real user data was accessed, exfiltrated, or retained. All identified vulnerabilities were responsibly disclosed to Telegram and affected developers prior to publication, and \HIGHLIGHT{Telegram subsequently introduced new secure-storage APIs and confirmed remediation of the official Wallet vulnerability.}

\noindent\textbf{Data minimization and storage.}
All experiments were conducted on dedicated research accounts with no personal or financial data. Storage dumps extracted during analysis were retained only for the duration of the vulnerability validation phase and stored on encrypted, access-controlled research infrastructure. No mnemonic phrases, private keys, or JWTs from real user accounts were collected at any point. The ground-truth evaluation dataset (Section~\ref{sect:tenet_eval}) was constructed entirely from storage generated by our own research accounts.

\noindent\textbf{Deletion procedures.}
All extracted storage artifacts were securely deleted upon completion of each experimental phase using cryptographic erasure. No data was transferred outside the research team or retained beyond the end of the validation period.

\noindent\textbf{Responsible disclosure timeline.}
We notified Telegram's security team and all affected Mini App developers upon completing our initial analysis, providing a detailed report with per-app findings and a working proof-of-concept. We granted a 90-day remediation window before public disclosure, in line with industry-standard coordinated vulnerability disclosure practice. Telegram acknowledged our report, introduced the SecureStorage and DeviceStorage APIs within the disclosure window, and confirmed remediation of the official Wallet vulnerability. Individual Mini App developers were contacted directly; developers who did not respond within 90 days were notified of the impending disclosure.

\noindent\textbf{TENET tool release.}
TENET is designed for legitimate security auditing of one's own devices and Mini App deployments. Its release will be accompanied by a usage policy restricting its use to authorized environments. The tool does not include any network exfiltration capability in its public release; the PoC exfiltration component described in Section~\ref{sect:vuln_analysis} was used solely for internal validation and is not distributed.

\bibliographystyle{plain}
\bibliography{ref}

\end{document}